\documentclass[%
aps,
twocolumn,
superscriptaddress,
amsmath,amssymb,
10pt
]{revtex4-2}
\usepackage{url}
\usepackage{graphicx}
\usepackage{subcaption}
\usepackage{placeins}
\usepackage{dcolumn}
\usepackage{bm}
\usepackage[colorlinks=true,linkcolor=blue,citecolor=blue]{hyperref}

\usepackage{overpic}
\usepackage{physics}
\usepackage{dsfont}
\usepackage{tikz,pgfplots}
\usepackage{enumerate}
\usepackage{bbold}
\usepackage{makecell}
\usepackage{array}
\makeatletter
\newcommand{\thickhline}{%
    \noalign {\ifnum 0=`}\fi \hrule height 1pt
    \futurelet \reserved@a \@xhline
}
\newcolumntype{"}{@{\hskip\tabcolsep\vrule width 1pt\hskip\tabcolsep}}
\makeatother
\usepackage{multirow}
\usepackage{textcomp,booktabs}
\usepackage{colortbl}
\usepackage[T1]{fontenc}
\definecolor{mygray}{gray}{0.9}
\definecolor{mypink}{rgb}{0.99,0.91,0.95}
\definecolor{mycyan}{cmyk}{0.3,0,0,0}

\newcommand{\1}{\mathbb{1}}

\definecolor{KB1}{rgb}{0.4,0.3,0.9}
\definecolor{KB2}{rgb}{0.9,0.3,0.9}

\definecolor{daxColor}{HTML}{900C3F}

\makeatletter
\newcommand*{\rom}[1]{\expandafter\@slowromancap\romannumeral #1@}
\makeatother

\usepackage{amsthm}
\newtheorem*{thm*}{Theorem}

\usepackage{tcolorbox}
\tcbuselibrary{theorems}

\newtcbtheorem[number within=section, use counter*=thm]{mythm}{Theorem}%
{colback=red!5,colframe=red!35!red,fonttitle=\bfseries}{thm}

\usepackage{tcolorbox}
\tcbuselibrary{theorems}

\newtcbtheorem[number within=section, use counter*=thm]{mylem}{Lemma}%
{colback=green!5,colframe=green!35!black,fonttitle=\bfseries}{lem}

\usepackage{tcolorbox}
\tcbuselibrary{theorems}

\newtcbtheorem[number within=section, use counter*=thm]{mydef}{Definition}%
{colback=yellow!5,colframe=yellow!70!black,fonttitle=\bfseries}{def}

\usepackage{tcolorbox}
\tcbuselibrary{theorems}

\newtcbtheorem[number within=section, use counter*=thm]{mycor}{Corollary}%
{colback=blue!5,colframe=blue!70!black,fonttitle=\bfseries}{cor}

\usepackage{tcolorbox}
\tcbuselibrary{theorems}

\newtcbtheorem[number within=section, use counter*=thm]{myprop}{Proposition}%
{colback=pink!5,colframe=pink!70!black,fonttitle=\bfseries}{prop}

\pgfplotsset{compat=1.17}

\begin{document}


\title{Nonlocal advantage of quantum imaginarity in Schwarzschild spacetime}

\author{Bing Yu}
\affiliation{School of Mathematics and Systems Science, Guangdong Polytechnic Normal University, Guangzhou 510665, China}

\author{Xiao-Yong Yang}
\affiliation{School of Mathematics and Statistics, Central China Normal University, Wuhan 430079, China}
\author{Xiaoli Hu}
\affiliation{School of Artificial Intelligence, Jianghan University, Wuhan 430056, China}

\author{Zhi-Xiang Jin}
\email{jzxjinzhixiang@126.com}
\affiliation{School of Computer Science and Technology, Dongguan University of Technology, Dongguan 523808, China}

\author{Xiaofen Huang}
\email{huangxf1206@163.com}
\affiliation{School of Mathematics and Statistics, Hainan Normal University, Haikou 571158, China}

\date{\today}
             
\begin{abstract} 
Black hole spacetimes provide a natural setting for quantum systems in curved spacetime, where effects such as Hawking radiation arise from event horizons. 
In this work, we investigate the impact of Hawking radiation on quantum imaginarity in Schwarzschild spacetime, focusing on nonlocal advantage of quantum imaginarity (NAQI) and assisted imaginarity distillation.
For NAQI, Hawking radiation induces opposite temperature dependence of the NAQI gap in the two regions, decreasing in the physically accessible region and increasing in the physically inaccessible region.
Consequently, the gap can cross zero only in the accessible region, leading to the disappearance of NAQI at sufficiently high temperature, while it remains negative in the inaccessible region.
For assisted imaginarity distillation, the fidelity decreases with temperature in the physically accessible region and increases in the physically inaccessible region, resulting in reduced and enhanced distillation capability, respectively.
These results reveal distinct operational behaviors of physically accessible and inaccessible regions under relativistic effects, providing insight into quantum imaginarity in curved spacetime.
\end{abstract}

\maketitle



\section{Introduction}
The complex-valued structure of quantum mechanics underlies a resource-theoretic distinction between real and non-real quantum states defined with respect to a fixed reference basis,
where it is identified as a resource known as imaginarity, introduced by Hickey and Gour~\cite{quantifying_hickey2018}. 
In this framework, states with real density matrices are regarded as free, and free operations are those that do not generate imaginarity.
Since its introduction, quantum imaginarity has been extensively studied from different perspectives.
It has been quantified through various measures, including norm-based, robustness-based, entropic-based, and fidelity-based approaches, among others~\cite{quantifying_hickey2018,operational_imaginarity_wu2021,measures_imaginarity_Gao2023,resource_imagianrity_wu2021,quantification_xue2021,Tsallis_xu2024,imaginarity_Gaussian_xu2023,real_operations_Streltsov2023,pure_imaginarity_du2025}.
Beyond quantification, imaginarity has been shown to provide operational advantages in quantum information processing tasks, notably in state and channel discrimination ~\cite{operational_imaginarity_wu2021,imaginarity_distributed_wu2024}.
In parallel, its interconvertible with other quantum resources, including entanglement, coherence and quantum discord, has also been investigated~\cite{imaginarity_entanglement_sun2025,imaginarity_coherence_xu2025,imaginarity_discord_jin2026}.
Despite these advances, the distribution of imaginarity in composite systems has also attracted significant interest, revealing nonlocal features such as the nonlocal advantage of quantum imaginarity~\cite{NAQIwei2024} and assisted imaginarity distillation~\cite{imaginarity_distributed_wu2024}.

Quantum information processing in relativistic settings has attracted considerable interest in recent years, particularly in curved spacetime scenarios such as black hole backgrounds.
In Schwarzschild spacetime, the presence of an event horizon gives rise to Hawking radiation, which effectively introduces thermal effects that can reshape the distribution of quantum correlations across different modes.
A substantial body of work has explored how such relativistic effects influence various quantum correlations, including entanglement, coherence, discord, as well as quantum nonlocality and  steering~\cite{entanglement_redistribution_wang2010,unveiling_entang_martin2010,wang2010projective,entang_distill_deng2011,multipartite_entanglement_xu2014,distinguishability_discord_shi2018,harvesting_tjoa2020,steering_monogamy_wu2022,GTriNonloczhang2023,teleportation_wu2023,quantumness_haddadi2024,genuinely_entanglement_wu2024,maximal_steered_coherence_du2024,physically_accessible_elghaayda2024,steering_coherence_wang2025,harvesting_entanglement_dubey2025,fully_entanled_fraction_mi2025,trip_coherencejoyia2025,generation_entanglement_liu2025,tripartite_steering_mi2026,entanglement_li2026,discord_xiong2026,ming2026nonlocal,EUR_yao2026,continuous_coherence_liu2026}.
It has been shown that Hawking radiation can significantly modify quantum correlations in the physically accessible region, while also inducing a redistribution of correlations between exterior and interior modes of the black hole spacetime.
Despite these extensive studies, the behavior of quantum imaginarity remains unknown in curved spacetime.
This naturally raises the question of how imaginarity is affected by Hawking radiation, which is essential for a more complete understanding of quantum resource theories in relativistic regimes.

In this work, we investigate the behavior of quantum imaginarity in Schwarzschild spacetime by focusing on two representative operational tasks, namely the nonlocal advantage of quantum imaginarity (NAQI) and assisted imaginarity distillation.
By analyzing the Hawking-induced transformation of quantum states, we study how imaginarity is distributed between physically accessible and inaccessible regions.
We find that the NAQI gap is governed by Hawking-induced effects that produce opposite responses in the two spacetime regions, decreasing in the physically accessible region and increasing in the physically inaccessible region.
As a result, the gap can cross zero in the accessible region for certain parameter regimes, while remaining negative in the inaccessible region.
This behavior indicates that NAQI emerges only within restricted regions of the parameter space.
In regimes where NAQI is present, the underlying states exhibit steerability.
We further examine assisted imaginarity distillation under the same Hawking-induced dynamics. 
The distillation fidelity displays opposite temperature-dependent behavior in the two regions, decreasing in the physically accessible region and increasing in the physically inaccessible region. 
This leads to a reduction of the assisted distillation capability in the accessible region and an enhancement in the inaccessible region, highlighting distinct operational responses of the two subsystems under relativistic effects.

The remainder of this paper is organized as follows.
Sec.~\ref{sec:Schwarzschild_background} presents the quantization of the Dirac field in Schwarzschild spacetime.
Sec.~\ref{sec:NAQI} introduces the framework of NAQI. 
In Sec.~\ref{sec:NAQI_Schwarzschild}, we analyze its behavior in Schwarzschild spacetime for two-qubit states.
Sec.~\ref{sec:AID} investigates assisted imaginarity distillation in Schwarzschild spacetime.
Finally, Sec.~\ref{sec:Conclusion} concludes the paper.
\section{Quantization of Dirac field in Schwarzschild black hole}
\label{sec:Schwarzschild_background}

The metric of the Schwarzschild black hole is given by
\begin{align}
ds^2=&-(1-\frac{2M}{r}) dt^2+(1-\frac{2M}{r})^{-1} dr^2\nonumber\\
&+r^2(d\theta^2
+\sin^2\theta d\varphi^2),
\end{align}
where $M$ denotes the mass of the black hole.
We adopt natural units $\hbar = G = c = k = 1$.
The massless Dirac equation~\cite{brill1957interaction} $[\gamma^a e_a^\mu(\partial_\mu+\Gamma_\mu)]\Phi=0$, where $\gamma^a$ represent the Dirac matrice, $e_a^\mu$ is the tetrad field and $\Gamma_\mu$ denotes the spin connection, can be  written explicitly as
\begin{align}
&\frac{-\gamma_0}{\sqrt{1-\frac{2M}{r}}}\frac{\partial \Phi}{\partial t}+\gamma_1\sqrt{1-\frac{2M}{r}}\bigg[\frac{\partial}{\partial r}+\frac{1}{r}+\frac{M}{2r(r-2M)} \bigg]\Phi\notag\\
&+\frac{\gamma_2}{r}(\frac{\partial}{\partial \theta}+\frac{\cot \theta}{2})\Phi+\frac{\gamma_3}{r\sin\theta}\frac{\partial\Phi}{\partial\varphi}=0.
\end{align}

To analyze the field behavior near the event horizon, it is convenient to introduce the tortoise coordinate $r_{*}=r+2M\ln\frac{r-2M}{2M}$, in terms of which the radial part of the wave equation becomes regular at the horizon.
Defining the retarded time $u = t - r_*$, one finds that the Dirac equation admits separable solutions. In the near-horizon limit, the dominant contributions correspond to outgoing modes of the form
\begin{align}
\Phi^+_{{\mathbf{k}},{\rm in}}&\sim \phi(r) e^{i\omega u},\label{outgoing_mode1}\\
\Phi^+_{{\mathbf{k}},{\rm out}}&\sim \phi(r) e^{-i\omega u},\label{outgoing_mode2}
\end{align}
where $\phi(r)$ is the four-component spinor.
For a massless field, the frequency and momentum satisfy the dispersion relation $|\mathbf{k}| = \omega$.
The field operator can then be quantized by expanding it in terms of these Schwarzschild modes,
\begin{align}\label{direc_field}
\Phi=&\int
d\mathbf{k}[\hat{a}^{\rm in}_{\mathbf{k}}\Phi^{+}_{{\mathbf{k}},\text{in}}
+\hat{b}^{\rm in \dagger}_{\mathbf{k}}
\Phi^{-}_{{\mathbf{k}},\text{in}}\nonumber\\ 
&+\hat{a}^{\rm out}_{\mathbf{k}}\Phi^{+}_{{\mathbf{k}},\text{out}}+\hat{b}^{\rm out \dagger}_{\mathbf{k}}\Phi^{-}_{{\mathbf{k}},\text{out}}],
\end{align}
where $\hat{a}^{\rm \eta}_{\mathbf{k}}$ and $\hat{b}^{\rm \eta\dagger}_{\mathbf{k}}$ with $\eta=(\mathrm{in}, \mathrm{out})$ are the fermion annihilation and creation operators for particles and antiparticles, respectively. 
The Schwarzschild vacuum $|0\rangle_S$ is defined by $\hat{a}^{\rm in}_{\mathbf{k}}|0\rangle_S=\hat{a}^{\rm out}_{\mathbf{k}}|0\rangle_S=0$.

Following the approach proposed by Damour and Ruffini~\cite{damour1976black}, a complete basis of global positive-energy modes can be constructed via analytic continuation of Eq.~(\ref{outgoing_mode1}) and Eq.~(\ref{outgoing_mode2}) across the horizon. This yields the Kruskal modes
\begin{align}
\Psi^+_{{\mathbf{k}},{\rm out}}
&=e^{-2\pi M\omega} \Phi^-_{{-\mathbf{k}},{\rm in}}+e^{2\pi M\omega}\Phi^+_{{\mathbf{k}},{\rm out}},\\
\Psi^+_{{\mathbf{k}},{\rm in}}
&=e^{-2\pi M\omega} \Phi^-_{{-\mathbf{k}},{\rm out}}+e^{2\pi M\omega}\Phi^+_{{\mathbf{k}},{\rm in}}.
\end{align}
Using these modes, the field can equivalently be expanded in the Kruskal basis,
\begin{align}\label{dirac_filed2}
\Phi=&\int
d\mathbf{k} [2\cosh(4\pi M\omega)]^{-\frac{1}{2}}
[\hat{c}^{\rm in}_{\mathbf{k}}\Psi^{+}_{{\mathbf{k}},\text{in}}
+\hat{d}^{\rm in\dagger}_{\mathbf{k}}
\Psi^{-}_{{\mathbf{k}},\text{in}}\nonumber\\ &+\hat{c}^{\rm out}_{\mathbf{k}}\Psi^{+}_{{\mathbf{k}},\text{out}}
+\hat{d}^{\rm out\dagger}_{\mathbf{k}}\Psi^{-}_{{\mathbf{k}},\text{out}}],
\end{align}
where $\hat{c}^{\eta}_{\mathbf{k}}$ and $\hat{d}^{\eta\dagger}_{\mathbf{k}}$ denote the annihilation and creation operators acting on the Kruskal vacuum.

By matching Eqs.~(\ref{direc_field}) and ~(\ref{dirac_filed2}), one obtains the Bogoliubov transformation~\cite{barnett2002methods} relating the Schwarzschild and Kruskal operators.
For the exterior region, this relation takes the form
\begin{align}
\hat{c}^{\rm out}_{\mathbf{k}}&=\frac{1}{\sqrt{e^{-8\pi M\omega}+1}}\hat{a}^{\rm out}_{\mathbf{k}}-\frac{1}{\sqrt{e^{8\pi M\omega}+1}}\hat{b}^{\rm in\dagger}_{\mathbf{k}},\\
\hat{c}^{\rm out\dagger}_{\mathbf{k}}&=\frac{1}{\sqrt{e^{-8\pi M\omega}+1}}\hat{a}^{\rm out\dagger}_{\mathbf{k}}-\frac{1}{\sqrt{e^{8\pi M\omega}+1}}\hat{b}^{\rm in}_{\mathbf{k}}.
\end{align}

Consequently, the Kruskal vacuum state $|0\rangle_K$ and excited state $|1\rangle_K$ can be expressed explicitly as \cite{wang2010projective}
\begin{align}\label{isometry}
|0\rangle_K=
&\frac{1}{\sqrt{e^{-\frac{\omega}{T}}+1}}|0\rangle_{\rm out} |0\rangle_{\rm in}
+\frac{1}{\sqrt{e^{\frac{\omega}{T}}+1}}|1\rangle_{\rm out} |1\rangle_{\rm in},\notag\\
|1\rangle_K=&|1\rangle_{\rm out} |0\rangle_{\rm in},
\end{align}
where $T=\frac{1}{8\pi M}$ is the Hawking temperature. Here, $\{|n\rangle_{\rm out}\}$ and $\{|n\rangle_{\rm in}\}$ denote the Schwarzschild number states for fermions outside the event horizon and antifermions inside the event horizon.

\section{nonlocal advandage of imaginarity}
\label{sec:NAQI}

The imaginarity of a quantum state $\rho$ with respect to a reference basis $\hat{\mathcal{B}}$ can be quantified by various measures.
In particular, the $l_1$-norm of imaginarity is defined as
\begin{align}
\mathcal{I}_{l_1,\hat{\mathcal{B}}}(\rho) = \sum_{i\neq j} |\mathrm{Im}(\rho_{ij})|,
\end{align}
where $\rho_{ij}$ are the matrix elements of $\rho$ in the basis $\hat{\mathcal{B}}$, and $\mathrm{Im}(\rho_{ij})$ denotes their imaginary part~\cite{quantifying_hickey2018}.
The relative entropy of imaginarity is defined as
\begin{align}
\mathcal{I}_{rel,\hat{\mathcal{B}}}(\rho) = S(Re(\rho)) - S(\rho),
\end{align}
where $Re(\rho)=\frac{\rho+\rho^t}{2}$ denotes the real part of $\rho$, with $t$ representing the transpose, and $S(\rho)$ is the von Neumann entropy~\cite{quantification_xue2021}.
In what follows, we collectively denote these measures by $\mathcal{I}_{q,\hat{\mathcal{B}}}(\rho)$ with $q\in\{l_1,rel\}$.

On the single-qubit level, imaginarity satisfies a complementarity relation across mutually unbiased bases(MUBs) $\mathcal{B} = \{\mathcal{B}_x\}_{x=1}^3$, yielding an upper bound on the total imaginarity.
Explicitly, $\sum_x\mathcal{I}_{q,\mathcal{B}_x}(\rho)\leq \mathcal{I}_q$, where $\mathcal{I}_{l_1}=\sqrt{5}$ and $\mathcal{I}_{rel}\approx 2.02685$~\cite{NAQIwei2024}.
We now turn to the bipartite setting and consider the imaginarity that can be generated on one subsystem via local measurements on the other.
This leads to the notion of the nonlocal advantage of quantum imaginarity (NAQI), introduced in Ref.~\cite{NAQIwei2024}.
Specifically, Alice and Bob share a bipartite two qubit state $\rho^{AB}$.
Alice performs three binary projective measurements $\mathcal{P}=\{P_x\}_{x=1}^3$,with projectors $\{P_{o|x}\}_{o=0,1}$.
Each measurement induces conditional states $\rho^B_{o|x}$ on Bob’s side with probabilities $p(o|x)$, forming ensembles $\mathcal{E}_x=\{p(o|x), \rho^B_{o|x}\}$.
To quantify Bob’s imaginarity across different measurement settings, we evaluate the imaginarity of each ensemble $\mathcal{E}_x$ with respect to the corresponding basis $\mathcal{B}_x$.
The explicit forms of Alice’s projective measurements $\mathcal{P}$ and the corresponding mutually unbiased bases $\mathcal{B}$ used to evaluate Bob’s imaginarity are given in Appendix A.

Optimizing over all projective measurement strategies $\mathcal{P}$ and basis choices $\mathcal{B}$ leads to the quantity
\begin{align}\label{optim}
\mathcal{N}^{A\rightarrow B}_{q}(\rho^{AB}) = \max_{\mathcal{P},\mathcal{B}} \sum_{x,o} p(o|x) \mathcal{I}_{q,\mathcal{B}_x}\left(\rho^B_{o|x}\right),
\end{align}
where $q$ specifies the chosen measure of imaginarity, which can be either the $l_1$ norm or the relative entropy.
The arrow indicates the direction from Alice’s measurements to the resource quantified on Bob’s side.
 
Building on the single-system bound, NAQI is defined as its violation,
\begin{align}\label{NAQIdef}
\mathcal{N}^{A\rightarrow B}_{q}(\rho^{AB}) > \mathcal{I}_q.   
\end{align}

\section{NAQI in Schwarzschild spacetime}
\label{sec:NAQI_Schwarzschild}
In the following, we investigate NAQI of two qubit states in Schwarzschild spacetime. 
Motivated by Eq.~(\ref{NAQIdef}), we introduce the NAQI gap
\begin{align}\label{Delta}
\Delta_q(\rho^{AB})=\mathcal{N}^{A\rightarrow B}_{q}(\rho^{AB})-\mathcal{I}_q,  
\end{align}
where $q\in\{l_1,rel\}$, and $\mathcal{N}^{A\rightarrow B}_{q}(\rho^{AB})$ is defined in Eq.~(\ref{optim}).
The condition $\Delta_q > 0$ certifies the presence of NAQI, whereas $\Delta_q \le 0$ implies its absence.

We consider a bipartite two-qubit Bell-diagonal mixed state shared between Alice and Bob,
\begin{align}\label{eg1}
\rho^{AB} = p|\phi^+\rangle\langle\phi^+| + (1-p)|\psi^+\rangle\langle\psi^+|,
\end{align}
where $|\phi^+\rangle = \frac{|00\rangle + |11\rangle}{\sqrt{2}}$, $|\psi^+\rangle = \frac{|01\rangle + |10\rangle}{\sqrt{2}}$, and $p\in[0,1]$.
For later convenience, we also express $\rho^{AB}$ in the Bloch representation, which facilitates the analytical and numerical calculations.
Alice is assumed to remain in the asymptotically flat region, while Bob is in the vicinity of  the event horizon of a Schwarzschild black hole. 
Using the isometric extension induced by the Bogoliubov transformation in Eq.~(\ref{isometry}), the state $\rho^{AB}$ is mapped to a tripartite state $\rho^{AB_{\text{out}}B_{\text{in}}}$, where $B_{\text{out}}$ and $B_{\text{in}}$ denote the fermionic  modes outside and inside the event horizon, respectively. 
Due to the causal disconnection between the exterior and interior regions, the interior mode $B_{\text{in}}$ is inaccessible to external observers.
Accordingly, the accessible subsystem consists of ${A, B_{\text{out}}}$, while $B_{\text{in}}$ is treated as inaccessible.
By tracing over $B_{\text{in}}$ and $B_{\text{out}}$, we obtaion the physically accessible state $\rho^{AB_{\text{out}}}$ and  physical inaccessible state $\rho^{AB_{\text{in}}}$, respectively, given by
\begin{align}\label{eq1_out}
\rho^{AB_{\text{out}}}=&\frac{1}{4}
\big(\1 \otimes \1- \frac{1}{e^{\frac{\omega}{T}}+1} \1\otimes \sigma_3+ \frac{1}{\sqrt{e^{-\frac{\omega}{T}}+1}} \sigma_1\otimes \sigma_1 \notag\\
&+ (1-2p)\frac{1}{\sqrt{e^{-\frac{\omega}{T}}+1}}\sigma_2 \otimes \sigma_2\notag\\
&+ (2p-1)\frac{1}{e^{-\frac{\omega}{T}}+1} \sigma_3 \otimes \sigma_3\big), 
\end{align}

\begin{align}\label{eg1_in}
\rho^{AB_{\text{in}}}=&\frac{1}{4}
\big(\1 \otimes \1+ \frac{1}{e^{-\frac{\omega}{T}}+1} \1\otimes \sigma_3+ \frac{1}{\sqrt{e^{\frac{\omega}{T}}+1}} \sigma_1\otimes \sigma_1 \notag\\
&- (1-2p)\frac{1}{\sqrt{e^{\frac{\omega}{T}}+1}}\sigma_2 \otimes \sigma_2\notag\\
&- (2p-1)\frac{1}{e^{\frac{\omega}{T}}+1} \sigma_3 \otimes \sigma_3\big), 
\end{align}
where $\1$ denotes the identity matrix and $\{\sigma_i\}_{i=1}^3$ are Pauli matrices.
The derivation is straightforward and is therefore deferred to Appendix~B.

For the numerical analysis, it is convenient to parametrize the fermionic Bogoliubov coefficients by $\delta$, defined via $\cos\delta=1/\sqrt{e^{-\frac{\omega}{T}}+1}$ and $\sin\delta=1/\sqrt{e^{\frac{\omega}{T}}+1}$, with $\omega=1$.
For positive Hawking temperature $T>0$, the physical Schwarzschild regime corresponds to $\delta\in[0,\pi/4)$, which we focus on. 

\begin{figure}[htbp]
\centering
\begin{subfigure}{0.49\columnwidth}
    \centering
    \caption{}\includegraphics[width=\linewidth]{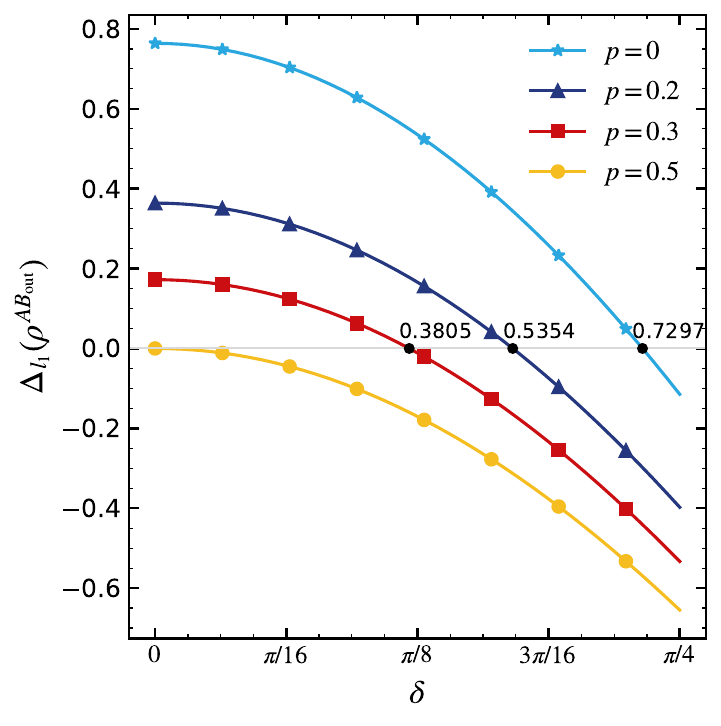}
\end{subfigure}
\hspace{-2pt}
\begin{subfigure}{0.49\columnwidth}
    \centering
    \caption{}\includegraphics[width=\linewidth]{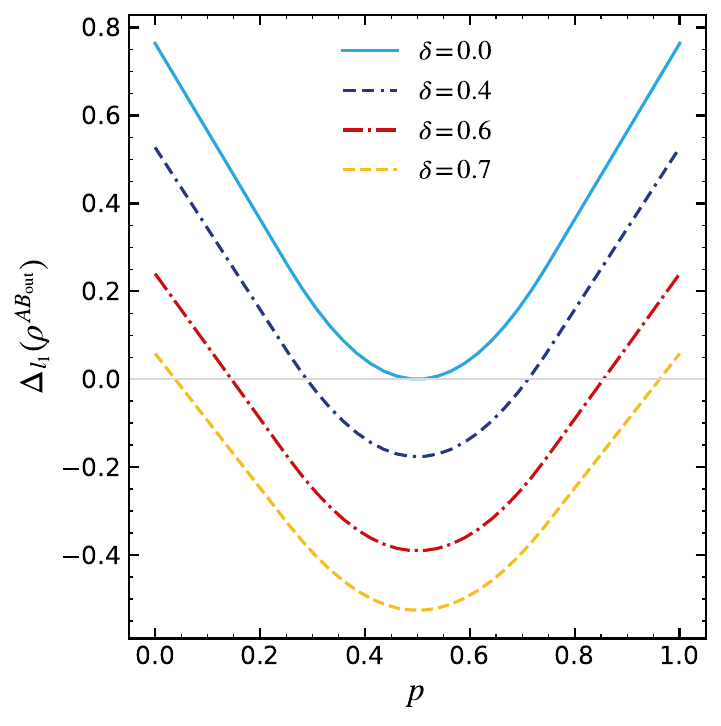}
\end{subfigure}
\vspace{-2pt}
\begin{subfigure}{0.49\columnwidth}
    \centering
    \caption{}\includegraphics[width=\linewidth]{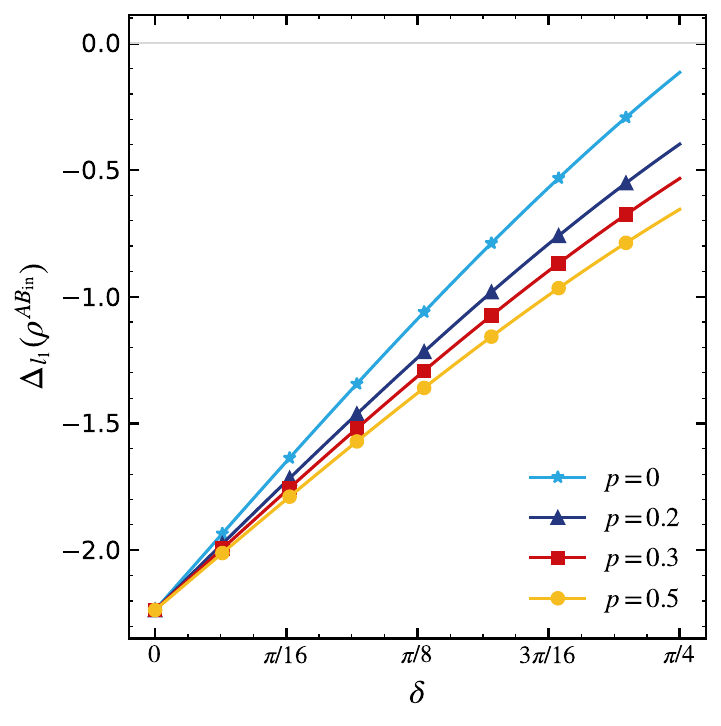}
\end{subfigure}
\hspace{-2pt}
\begin{subfigure}{0.49\columnwidth}
    \centering
        \caption{}
        \includegraphics[width=\linewidth]{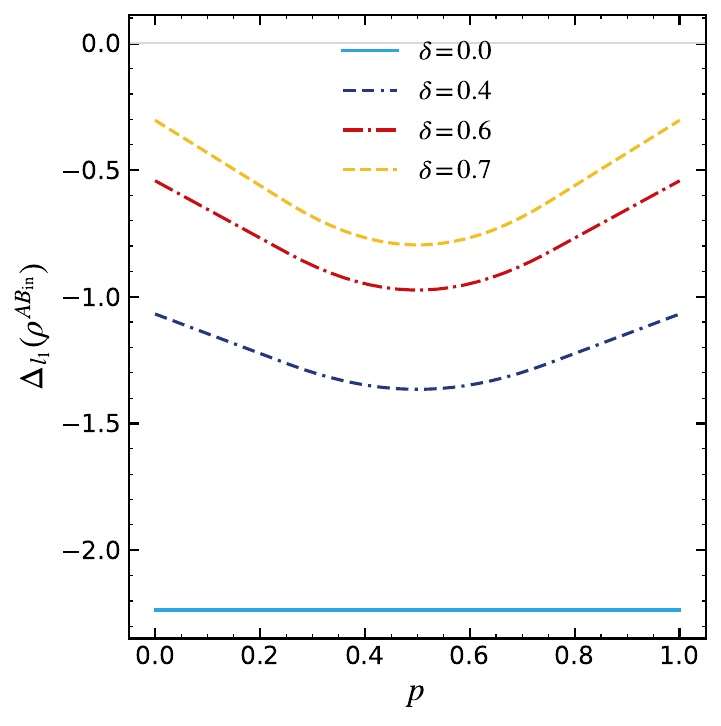}
\end{subfigure}
{
\captionsetup{justification=raggedright,singlelinecheck=false}
\caption{NAQI gap $\Delta_{l_1}$ for $\rho^{AB_{\text{out}}}$ and $\rho^{AB_{\text{in}}}$.
Panels $(a)$ and $(b)$ correspond to $\rho^{AB_{\text{out}}}$, while panels (c) and (d) correspond to $\rho^{AB_{\text{in}}}$.
Panels $(a)$ and $(c)$ show $\Delta_{l_1}$ as a function of $\delta$ for fixed $p=0,0.2,0.3,0.5$, with $\delta\in[0,\pi/4)$.
Panels $(b)$ and $(d)$ display $\Delta_{l_1}$ as a function of $p$ for fixed $\delta=0,0.4,0.6,0.7$, with $p\in[0,1]$. 
The horizontal gray solid line indicates $\Delta_{l_1}=0$.
The black dots mark the critical values of $\delta$ at which $\Delta_{l_1}$ becomes zero.} \label{l1eg1}}
\end{figure}

Fig.~\ref{l1eg1} presents the behavior of $\Delta_{l_1}(\rho^{AB_{\text{out}}})$ and $\Delta_{l_1}(\rho^{AB_{\text{in}}})$ as functions of $\delta$ and $p$.
A symmetry under $p \leftrightarrow 1-p$ is observed at fixed $\delta$, as shown in panels (b) and (d), allowing the analysis to be restricted to $p \leq 1/2$.

For the physical accessible state $\rho^{AB_{\text{out}}}$ (panels (a) and (b)), $\Delta_{l_1}$ decreases monotonically with increasing $\delta$ at fixed $p$.
In particular, $\Delta_{l_1}$ remains positive within a finite interval of $\delta$, whose extent depends on $p$.
Specifically, the positivity region is given by $\delta \lesssim 0.7297$ for $p=0$, $\delta \lesssim 0.5354$ for $p=0.2$, and $\delta \lesssim 0.3805$ for $p=0.3$, as illustrated in panel (a).
Since $\Delta_{q}>0$ provides a sufficient witness for imaginarity-based steerability~\cite{NAQIwei2024}. 
This behavior indicates a progressive shrinking of the witness region for imaginarity-based steerability under the influence of Hawking radiation.
In particular, for $p=0.5$, $\Delta_{l_1}$ remains non-positive for all $\delta$, implying the absence of NAQI at any temperature. 
In addition, increasing $p$ at fixed $\delta$ further reduces $\Delta_{l_1}$, indicating that the initial-state mixing and Hawking-induced effects act cooperatively in degrading the underlying imaginarity resource. 
At $\delta=0$, $\rho^{AB_{\text{out}}}$ reduces to the initial state $\rho^{AB}$, and the corresponding curve in panal (b) reproduces the result reported in Ref.~\cite{NAQIwei2024}.

For the physical inaccessible state $\rho^{AB_{\text{in}}}$ (panels (c) and (d)), $\Delta_{l_1}$ increases monotonically with $\delta$ but remains strictly negative throughout the entire parameter regime. 
Therefore, NAQI is absent for $\rho^{AB_{\text{in}}}$ at all temperatures.

Taken together, these results reveal a clear contrast between the exterior and interior sectors of Bob’s mode, Hawking radiation degrades and eventually destroys NAQI in the accessible mode, while the physical inacceible mode remains non-NAQI for all parameters.

\begin{figure}[htbp]
\centering
\begin{subfigure}{0.49\columnwidth}
    \centering
    \caption{}\includegraphics[width=\linewidth]{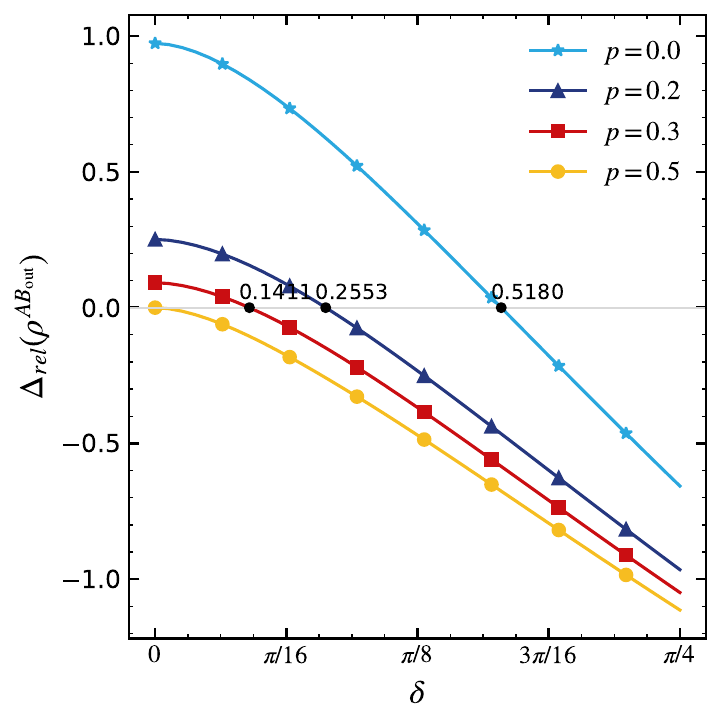}   
\end{subfigure}
\hspace{-2pt}
\begin{subfigure}{0.49\columnwidth}
    \centering
    \caption{}\includegraphics[width=\linewidth]{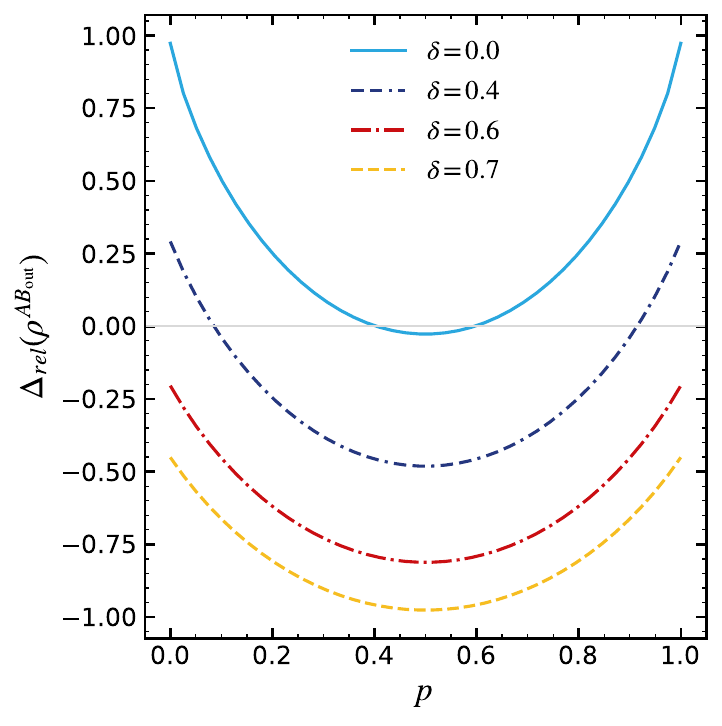}
\end{subfigure}
\vspace{-2pt}
\begin{subfigure}{0.49\columnwidth}
    \centering
    \caption{}\includegraphics[width=\linewidth]{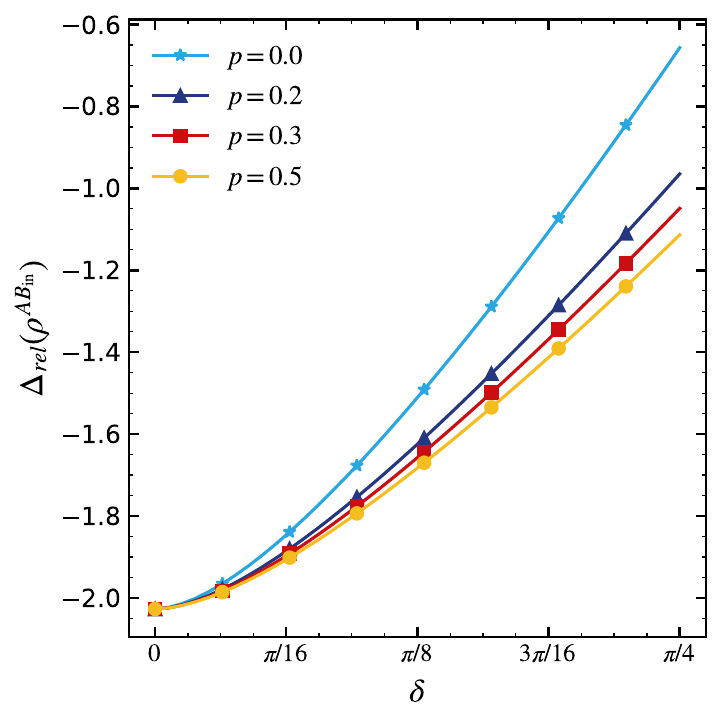}
\end{subfigure}
\hspace{-2pt}
\begin{subfigure}{0.49\columnwidth}
    \centering
        \caption{}
        \includegraphics[width=\linewidth]{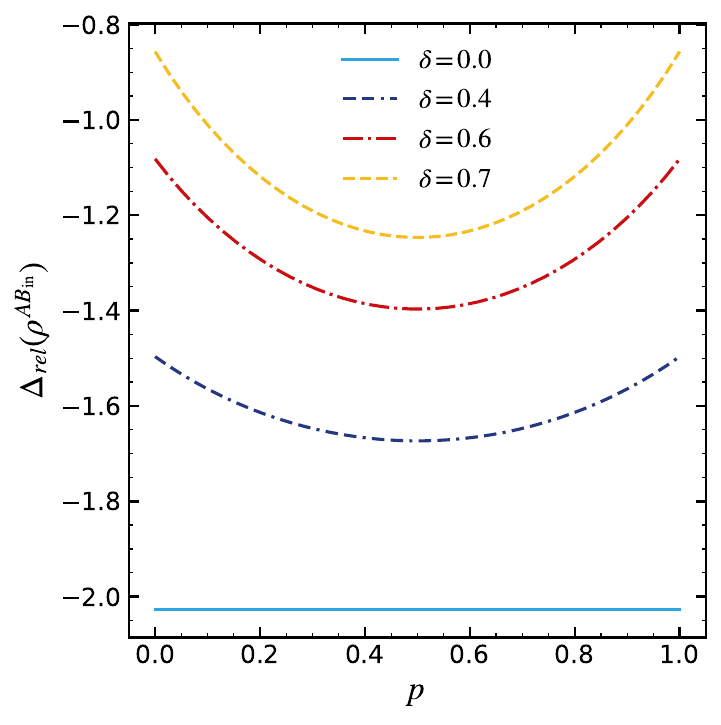}
\end{subfigure}
{
\captionsetup{justification=raggedright,singlelinecheck=false}
\caption{NAQI gap $\Delta_{rel}$ for $\rho^{AB_{\text{out}}}$ and $\rho^{AB_{\text{in}}}$.
Panels $(a)$ and $(b)$ correspond to $\rho^{AB_{\text{out}}}$, while panels (c) and (d) correspond to $\rho^{AB_{\text{in}}}$.
Panels (a) and (c) show $\Delta_{rel}$ as a function of $\delta$ for fixed $p=0,0.2,0.3,0.5$, with $\delta\in[0,\pi/4)$. 
Panels (b) and (d) show $\Delta_{rel}$ as a function of $p$ for fixed $\delta=0,0.4,0.6,0.7$, with $p\in[0,1]$. 
The horizontal gray line indicates $\Delta_{rel}=0$.
The black dots mark the critical values of $\delta$ at which $\Delta_{rel}$ reaches zero.}
\label{releg1}}
\end{figure}

We now move on to Fig.~\ref{releg1}, which presents the NAQI gap based on the relative entropy of imaginarity, denoted by $\Delta_{rel}(\rho^{AB_{\text{out}}})$ and $\Delta_{rel}(\rho^{AB_{\text{in}}})$. 
For the physically accessible state $\rho^{AB_{\text{out}}}$, $\Delta_{rel}$ decreases monotonically with increasing $\delta$ at fixed $p$, exhibiting a behavior qualitatively consistent with that observed in Fig.~\ref{l1eg1}.
For the physically inaccessible state $\rho^{AB_{\text{in}}}$, $\Delta_{rel}$ increases monotonically with $\delta$ at fixed $p$, similar to $\Delta_{l_1}$ in Fig.~\ref{l1eg1}.
Nevertheless, it remains in the negative regime throughout the entire parameter range.

These results consistently demonstrate a Hawking-induced opposite monotonic response in the exterior and interior sectors, which is stable across different imaginarity quantifiers. 

As a second example, we consider a two-qubit Werner state initially shared between Alice and Bob
\begin{align}\label{eg2}
\rho^{AB}_{W}=p |\phi^+\rangle \langle \phi^+| + \frac{1-p}{4} \1 \otimes \1
\end{align}
with $p\in[0,1]$.
The same Schwarzschild transformation in Eq.~(\ref{isometry}) is applied. 
The resulting reduced states in the exterior and interior regions are obtained analogously as
\begin{align}
\rho^{AB_{\text{out}}}_{W}= &\frac{1}{4}
\big(\1 \otimes \1- \frac{1}{e^{\frac{\omega}{T}}+1} \1\otimes \sigma_3+ p\frac{1}{\sqrt{e^{-\frac{\omega}{T}}+1}} \sigma_1\otimes \sigma_1 \notag\\
&-p\frac{1}{\sqrt{e^{-\frac{\omega}{T}}+1}}\sigma_2 \otimes \sigma_2+ p\frac{1}{e^{-\frac{\omega}{T}}+1} \sigma_3 \otimes \sigma_3\big).   
\end{align}

\begin{align}
\rho^{AB_{\text{in}}}_W=&\frac{1}{4}
\big(\1 \otimes \1+ \frac{1}{e^{-\frac{\omega}{T}}+1} \1\otimes \sigma_3+ p\frac{1}{\sqrt{e^{\frac{\omega}{T}}+1}} \sigma_1\otimes \sigma_1 \notag\\
&+p\frac{1}{\sqrt{e^{\frac{\omega}{T}}+1}}\sigma_2 \otimes \sigma_2- p\frac{1}{e^{\frac{\omega}{T}}+1} \sigma_3 \otimes \sigma_3\big). 
\end{align}

\begin{figure}[htbp]
\centering
\begin{subfigure}{0.49\columnwidth}
    \centering
    \caption{}\includegraphics[width=\linewidth]{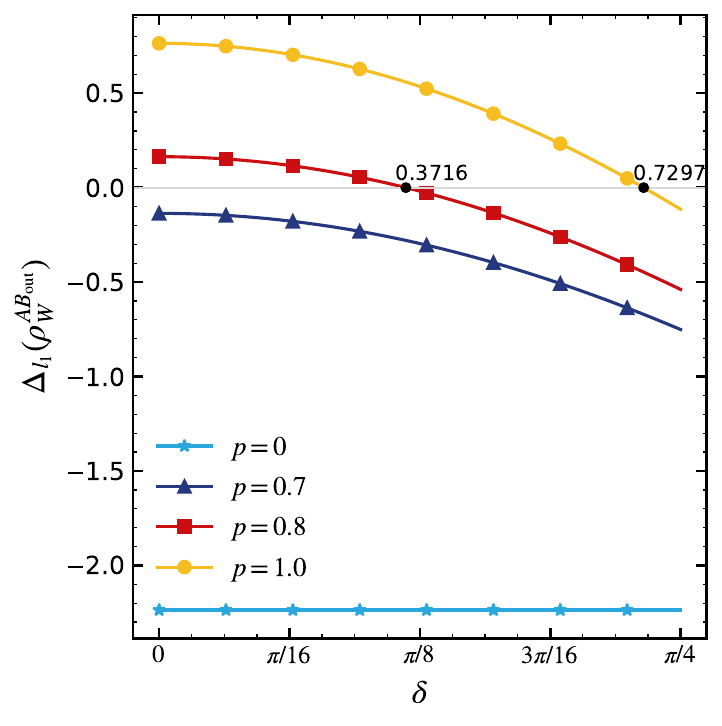}    
\end{subfigure}
\hspace{-2pt}
\begin{subfigure}{0.49\columnwidth}
    \centering
    \caption{}\includegraphics[width=\linewidth]{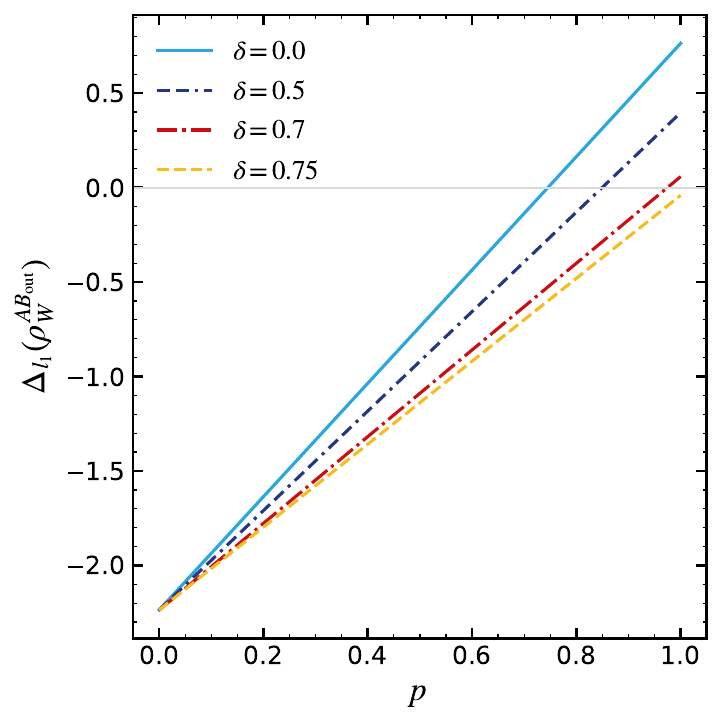}
\end{subfigure}
\vspace{-2pt}
\begin{subfigure}{0.49\columnwidth}
    \centering
    \caption{}\includegraphics[width=\linewidth]{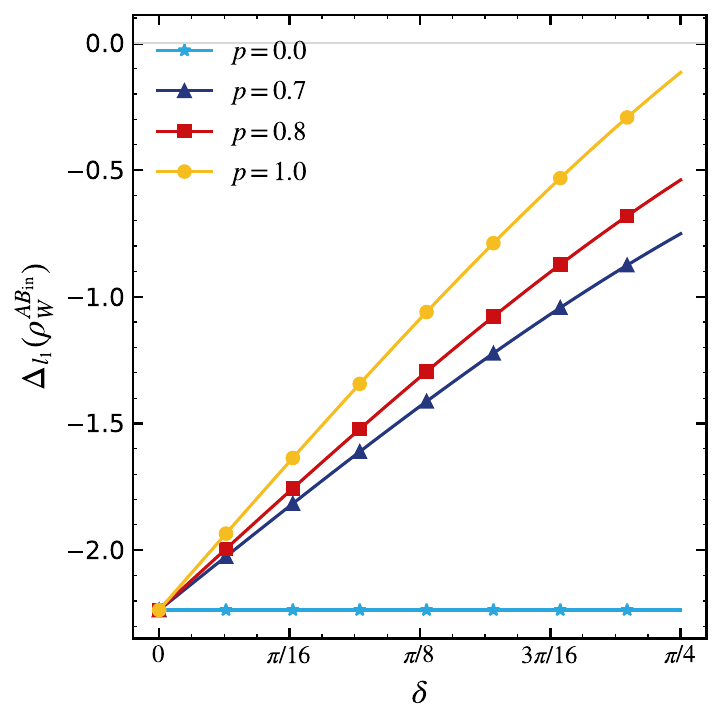}
\end{subfigure}
\hspace{-2pt}
\begin{subfigure}{0.49\columnwidth}
    \centering
        \caption{}
        \includegraphics[width=\linewidth]{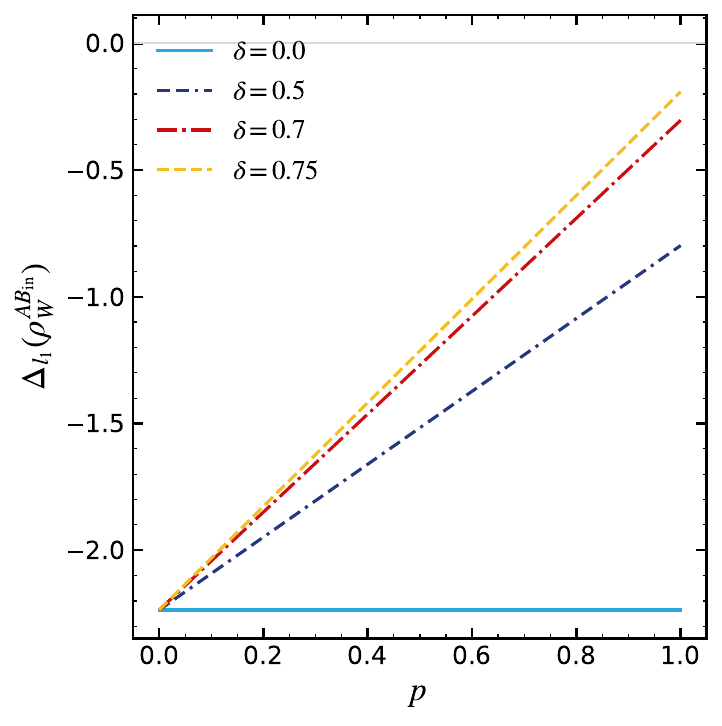}
\end{subfigure}
{
\captionsetup{justification=raggedright,singlelinecheck=false}
\caption{NAQI gap $\Delta_{l_1}$ for $\rho_{W}^{AB_{\mathrm{out}}}$ and $\rho_{W}^{AB_{\mathrm{in}}}$. 
Panels (a) and (c) show $\Delta_{l_1}$ as a function of $\delta$ for fixed $p=0,0.7,0.8,1.0$, with $\delta\in[0,\pi/4)$. 
Panels (b) and (d) show $\Delta_{l_1}$ as a function of $p$ for fixed $\delta=0,0.5,0.7,0.75$, with $p\in[0,1]$. 
The horizontal gray line indicates $\Delta_{l_1}=0$, and the black dots mark the corresponding critical values of $\delta$ at which $\Delta_{l_1}$ reaches zero. }
\label{l1eg2}}
\end{figure}

\begin{figure}[htbp]
\centering
\begin{subfigure}{0.49\columnwidth}
    \centering
    \caption{}\includegraphics[width=\linewidth]{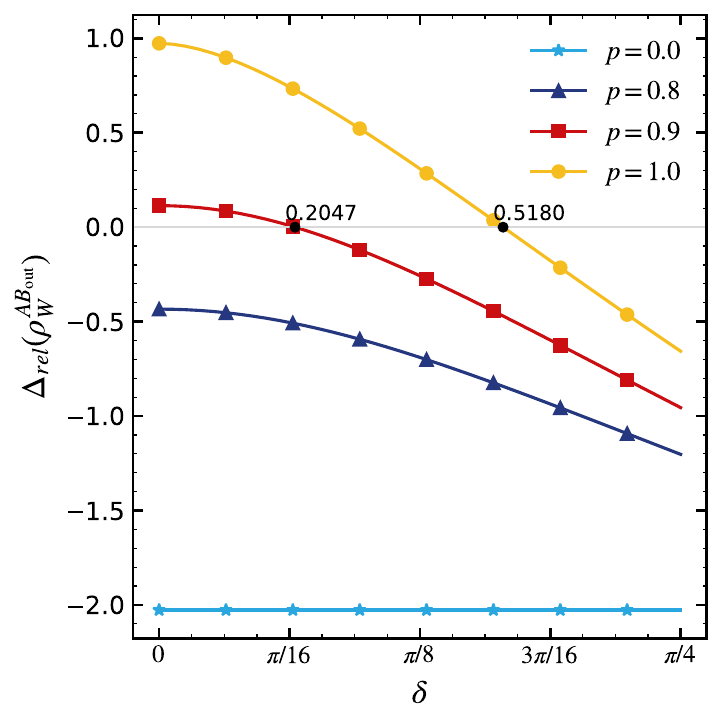}    
\end{subfigure}
\hspace{-2pt}
\begin{subfigure}{0.49\columnwidth}
    \centering
    \caption{}\includegraphics[width=\linewidth]{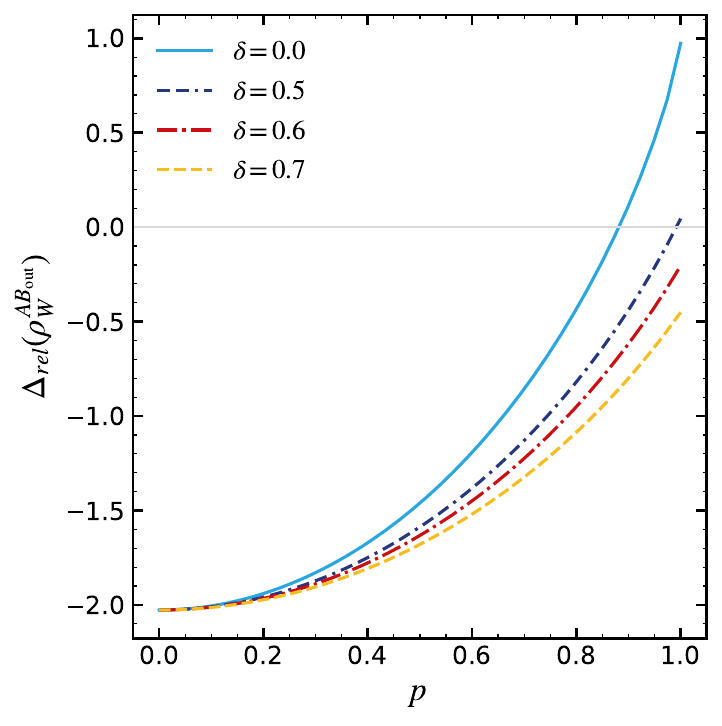}
\end{subfigure}
\vspace{-2pt}
\begin{subfigure}{0.49\columnwidth}
    \centering
    \caption{}\includegraphics[width=\linewidth]{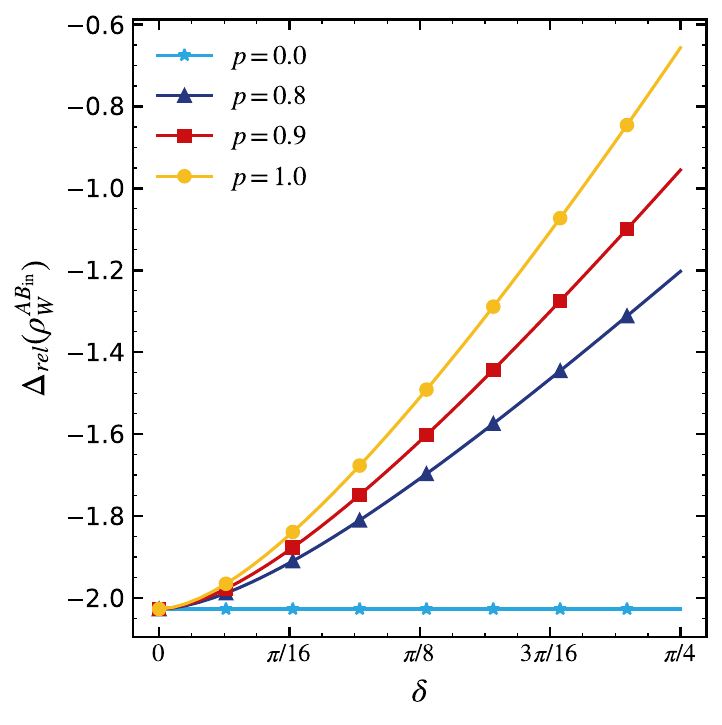}
\end{subfigure}
\hspace{-2pt}
\begin{subfigure}{0.49\columnwidth}
    \centering
        \caption{}
        \includegraphics[width=\linewidth]{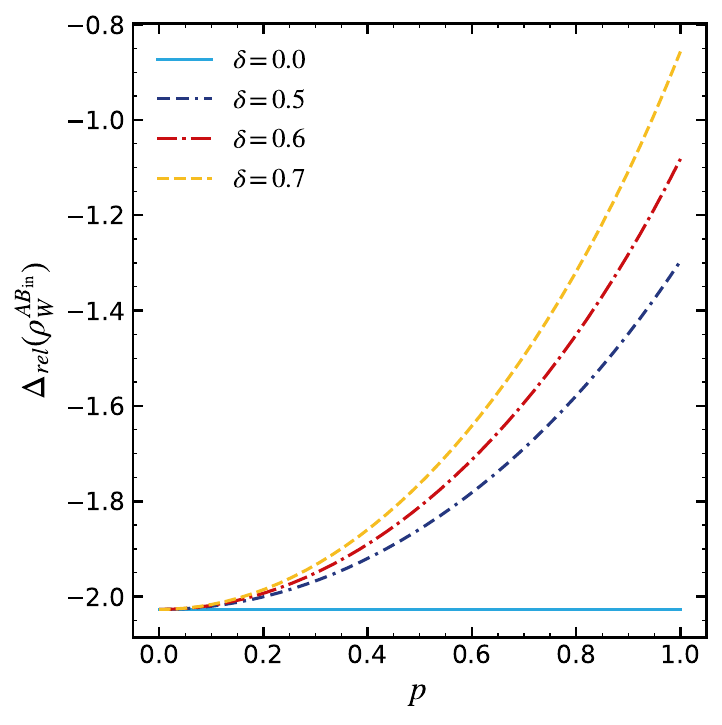}
\end{subfigure}
{\captionsetup{justification=raggedright,singlelinecheck=false}
\caption{NAQI gap $\Delta_{rel}$ for $\rho_{W}^{AB_{\mathrm{out}}}$ and $\rho_{W}^{AB_{\mathrm{in}}}$. 
Panels (a) and (c) show $\Delta_{rel}$ as a function of $\delta$ for fixed $p=0,0.8,0.9,1.0$, with $\delta\in[0,\pi/4)$. 
Panels (b) and (d) show $\Delta_{rel}$ as a function of $p$ for fixed  $\delta=0,0.5,0.6,0.7$, with $p\in[0,1]$. 
The horizontal gray line indicates $\Delta_{rel}=0$, and the black dots mark the corresponding critical values of $\delta$ at which $\Delta_{rel}$ reaches zero.
\label{releg2}}}
\end{figure}

We analyze the NAQI gaps $\Delta_{q}(\rho^{AB_{\text{out}}}_W)$ and $\Delta_{q}(\rho^{AB_{\text{in}}}_W)$ as functions of the relevant parameters.
In contrast to the first example, Fig.~\ref{l1eg2} and Fig.~\ref{releg2} reveal a qualitatively different dependence on $p$, where the NAQI gap exhibits a monotonic increase with $p$ at fixed $\delta$.
This contrasts with the previously observed nontrivial structure in the first example, indicating that the choice of initial state can significantly alter the parameter sensitivity of imaginarity under Hawking evolution.
For fixed $p$, the physically accessible and inaccessible sectors again display opposite monotonic responses with respect to $\delta$, consistent with the similar monotonic behavior observed in the previous example.
\section{Assisted imaginarity distillation in Schwarzschild spacetime}
\label{sec:AID}
We briefly recall the assisted imaginarity distillation protocol for bipartite quantum states 
$\rho^{AB}$ in Ref.~\cite{imaginarity_distributed_wu2024}, and study its extension to Schwarzschild spacetime.
This protocol is structurally analogous to the NAQI framework, in which Alice’s measurement induces a conditional ensemble on Bob’s subsystem. 
Alice performs a general positive operator-valued measure (POVM) $\{M_y\}_y$, yielding outcome $y$ with probability $p_y$, and Bob obtains the corresponding conditional state $\rho_y^B$.
The essential difference from the NAQI case is that Bob is additionally allowed to perform real operations $\Lambda_y$, aiming to distill the target maximally imaginary state $|\hat{+}\rangle=\frac{|0\rangle+\mathrm{i}|1\rangle}{\sqrt{2}}$. 
Here, the imaginarity is evaluated with respect to the computational basis, in contrast to the NAQI scenario where mutually unbiased bases are considered.
The performance of this  Alice-assisted protocol is quantified by the maximal achievable fidelity between Bob’s final state and the target state, optimized over all POVMs on Alice’s side and all real operations on Bob’s side.
This quantity, referred to as the assisted fidelity of imaginarity, admits a closed-form expression for any two-qubit state \cite{imaginarity_distributed_wu2024}
\begin{align}
    F_{d}(\rho^{AB})=\max_{M_y,\Lambda_y}\sum_y p_yF(\Lambda_y(\rho_y^B),|\hat{+}\rangle\langle\hat{+}|),
\end{align}
where $F(\rho,\sigma)=(\Tr\sqrt{\sqrt\sigma\rho\sqrt\sigma})^2$ denotes  quantum fidelity.
Importantly, we directly employ the analytical result of Ref.~\cite{imaginarity_distributed_wu2024}, which states that for any two-qubit state,
\begin{align}\label{optimalF}
 F_{d}(\rho^{AB})=\frac{1}{2}(1+\max\{|b_2|,|\vec{\mathrm{v}}|\}),
\end{align}
where $b_2=\Tr[\rho^{AB}(\1\otimes\sigma_2)]$, the vector $\vec{\mathrm{v}}=\{E_{12},E_{22},E_{32}\}$ with $E_{ij}=\operatorname{Tr}[\rho^{AB} (\sigma_i \otimes \sigma_j) ]$.

We now apply this framework to Schwarzschild spacetime.
We consider the same two-qubit state as in Eq.~(\ref{eg1}), with Alice in the asymptotically flat region and Bob near the event horizon.
The corresponding reduced states, $\rho^{AB_{\text{out}}}$ and $\rho^{AB_{\text{in}}}$ in Eq.~(\ref{eq1_out}) and Eq.~(\ref{eg1_in}), are obtained as before. 
We then evaluate the assisted fidelity of imaginarity for these states, namely $F_{d}(\rho^{AB_{\text{out}}})$ and $F_{d}(\rho^{AB_{\text{in}}})$.
According to Eq.~(\ref{optimalF}), we obtain
\begin{align}
 F_{d}(\rho^{AB_{\text{out}}})=\frac{1}{2}(1+|1-2p|\frac{1}{\sqrt{e^{-\frac{\omega}{T}}+1}}), 
\end{align}
and
\begin{align}\label{F_eg1_in_f}
F_{d}(\rho^{AB_{\text{in}}})=\frac{1}{2}(1+|1-2p| \frac{1}{\sqrt{e^{\frac{\omega}{T}}+1}}).
\end{align}

\begin{figure}[htbp]
\centering
\begin{subfigure}{0.49\columnwidth}
    \centering
    \caption{}\includegraphics[width=\linewidth]{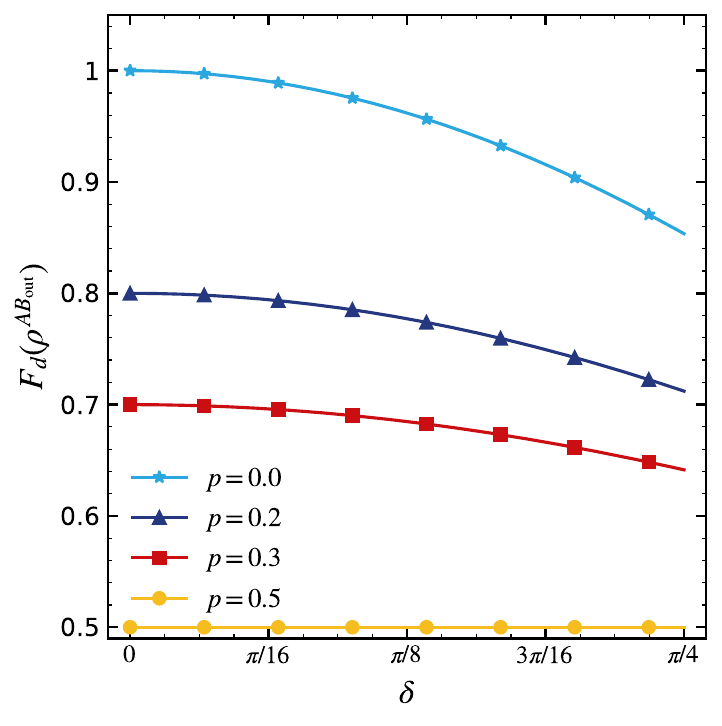}    
\end{subfigure}
\hspace{-2pt}
\begin{subfigure}{0.49\columnwidth}
    \centering
    \caption{}\includegraphics[width=\linewidth]{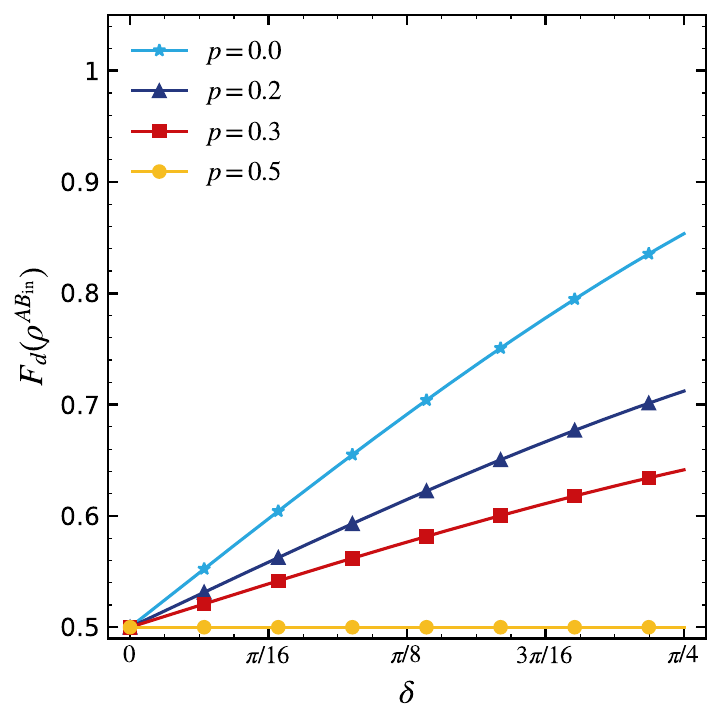}
\end{subfigure}
\vspace{-2pt}
\begin{subfigure}{0.49\columnwidth}
    \centering
    \caption{}\includegraphics[width=\linewidth]{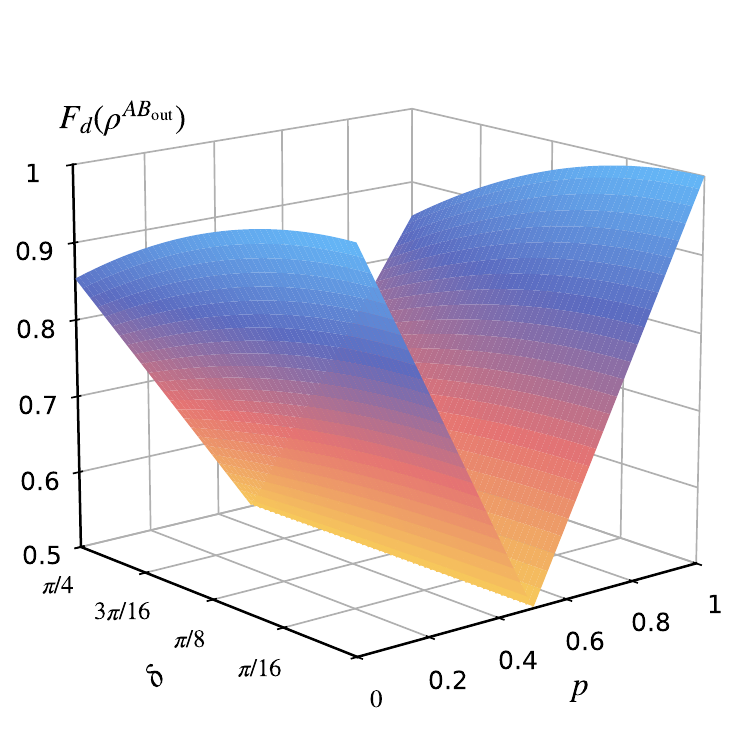}
\end{subfigure}
\hspace{-2pt}
\begin{subfigure}{0.49\columnwidth}
    \centering
        \caption{}
        \includegraphics[width=\linewidth]{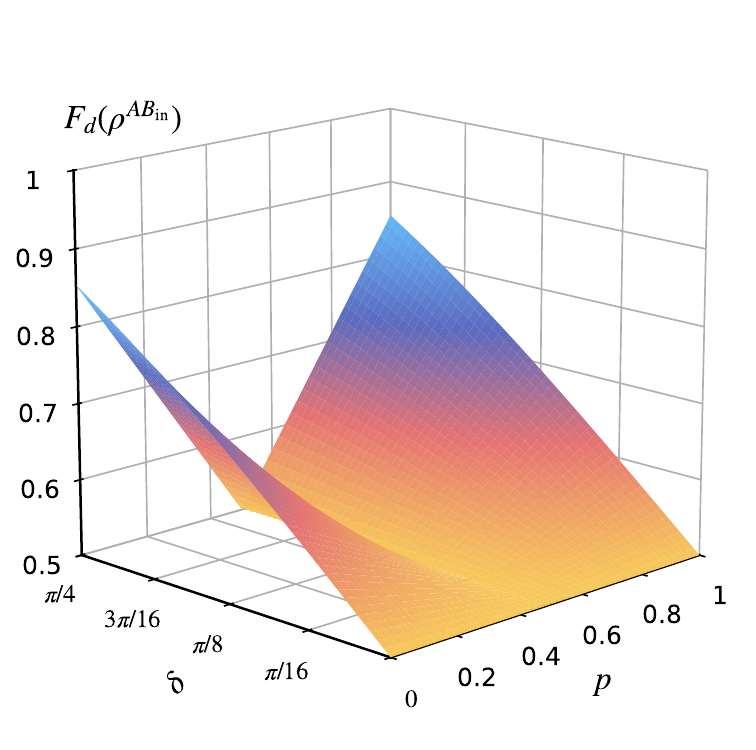}
\end{subfigure}
{\captionsetup{justification=raggedright,singlelinecheck=false}
\caption{
Assisted imaginarity fidelity $F_{d}(\rho^{AB_{\text{out}}})$ and $F_{d}(\rho^{AB_{\text{in}}})$. 
Panels $(a)$ and $(b)$ show the fidelity as a function of $\delta$ for $p=0,0.2,0.3,0.5$.
Panels $(c)$ and $(d)$ present the corresponding quantities over the parameter space $(\delta, p)$.
}
\label{F_eg1}}
\end{figure}

Fig.~\ref{F_eg1} shows the assisted imaginarity fidelity $F_{d}(\rho^{AB_{\text{out}}})$ and $F_{d}(\rho^{AB_{\text{in}}})$  as functions of $\delta$ and $p$.
As displayed in panels $(b)$ and $(d)$, both quantities exhibit a symmetry under $p \leftrightarrow 1-p$ at fixed $\delta$, which originates from the vanishing of $|1-2p|$ at $p=0.5$.
This symmetry leads to a nonmonotonic dependence on $p$.
For generic $\delta$, the fidelity first decreases and then increases as $p$ varies from $0$ to $1$. 
For the special case $\delta=0$, corresponding to zero temperature, one has $F_{d}(\rho^{AB_{\text{in}}})=1/2$ for all $p$.

For the physically accessible state $\rho^{AB_{\text{out}}}$, $F_{d}(\rho^{AB_{\text{out}}})$ decreases monotonically with increasing $\delta$ at fixed $p$, indicating that the Hawking effect suppresses the efficiency of imaginarity distillation.  
 
At the symmetric point $p=0.5$, one has 
$F_d(\rho^{AB_{\text{out}}}) = F_d(\rho^{AB_{\text{in}}}) = 1/2$ independently of $\delta$, 
showing that the distillation capability is insensitive to the Hawking temperature in this case.

In contrast, for the physical inaccessible state $\rho^{AB_{\text{in}}}$,  $F_{d}(\rho^{AB_{\text{out}}})$ exhibits the opposite behavior and increases monotonically with $\delta$ at fixed $p$, indicating an enhancement of distillation efficiency induced by the Hawking effect. 
The maximal value is bounded by $F_{d} \leq \frac{1}{2}(1+1/\sqrt{2}) \approx 0.8536$, as illustrated in panel $(b)$. 
In particular, for $p=0$, the fidelity approaches this upper bound as $\delta \to \pi/4^{-}$, i.e., in the infinite-temperature limit, as follows directly from Eq.~(\ref{F_eg1_in_f}).

For Werner state in Eq.~(\ref{eg2}), assisted fidelity of imaginarity for $\rho^{AB_{\text{out}}}$ and $\rho^{AB_{\text{in}}}$ is given by 
\begin{align}
 F_{d}(\rho_W^{AB_{\text{out}}})=\frac{1}{2}(1+p\frac{1}{\sqrt{e^{-\frac{\omega}{T}}+1}}), 
\end{align}
and
\begin{align}
F_{d}(\rho_W^{AB_{\text{in}}})=\frac{1}{2}(1+p \frac{1}{\sqrt{e^{\frac{\omega}{T}}+1}}).
\end{align}

\begin{figure}[htbp]
\centering
\begin{subfigure}{0.49\columnwidth}
    \centering
    \caption{}\includegraphics[width=\linewidth]{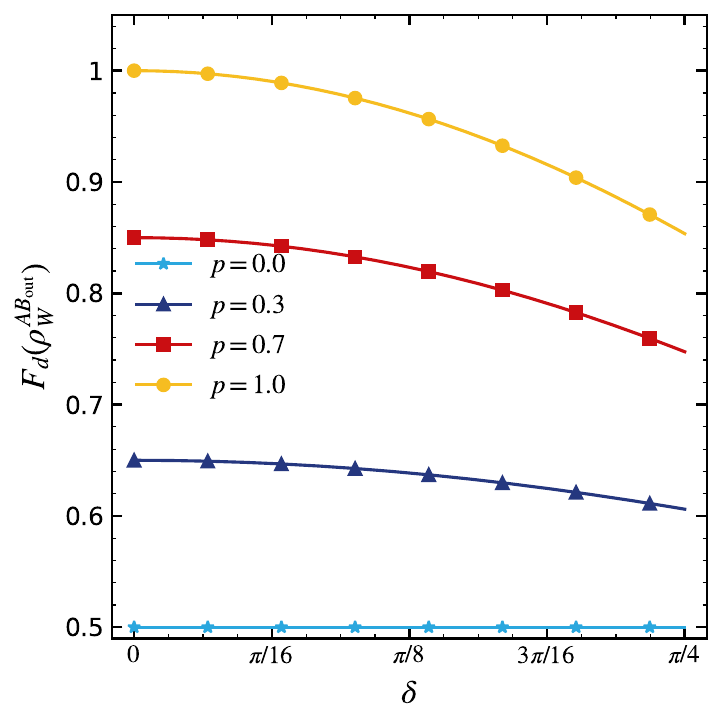}    
\end{subfigure}
\hspace{-2pt}
\begin{subfigure}{0.49\columnwidth}
    \centering
    \caption{}\includegraphics[width=\linewidth]{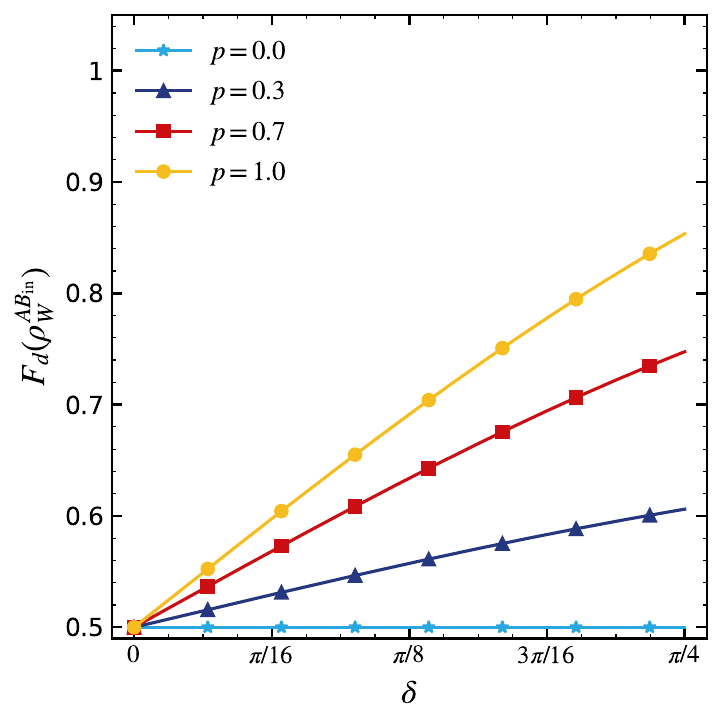}
\end{subfigure}
\vspace{-2pt}
\begin{subfigure}{0.49\columnwidth}
    \centering
    \caption{}\includegraphics[width=\linewidth]{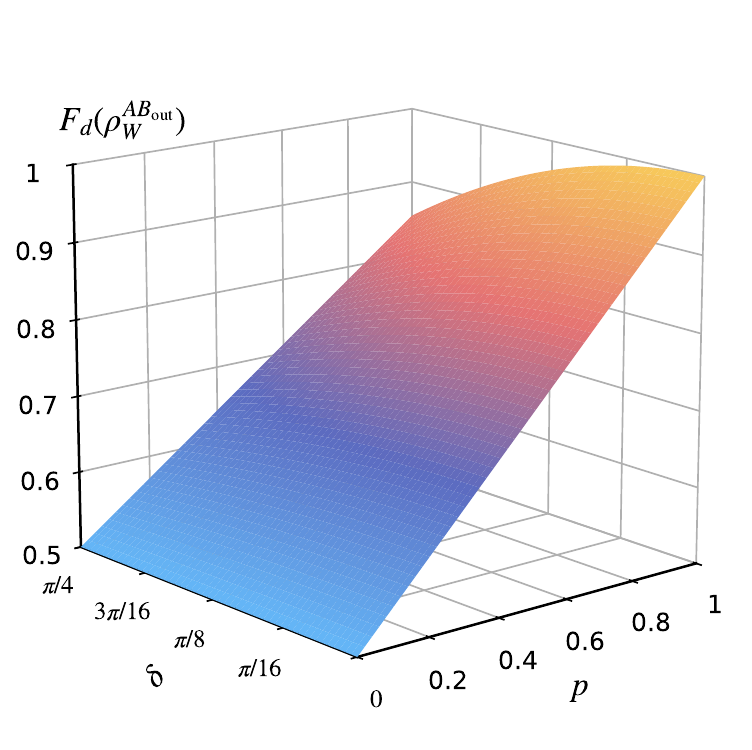}
\end{subfigure}
\hspace{-2pt}
\begin{subfigure}{0.49\columnwidth}
    \centering
        \caption{}
        \includegraphics[width=\linewidth]{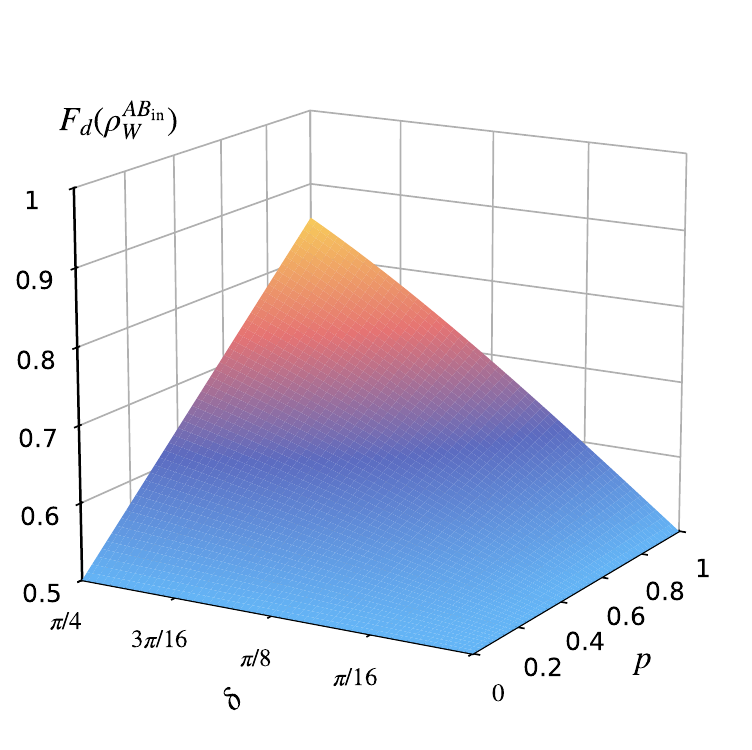}
\end{subfigure}
{\captionsetup{justification=raggedright,singlelinecheck=false}
\caption{Assisted imaginarity fidelity $F_{d}(\rho_W^{AB_{\text{out}}})$ and $F_{d}(\rho_W^{AB_{\text{in}}})$.
Panels $(a)$ and $(b)$ display the behavior of the quantity as $\delta$ varies for fixed $p=0,0.3,0.7,1$. 
Panels $(c)$ and $(d)$ show its variation across the parameter space $(\delta, p)$.
\label{F_eg2}}}
\end{figure}

For the Werner state, the behavior shown in Fig.~\ref{F_eg2} differs qualitatively from that in the previous example.
In particular, the dependence on $p$ becomes monotonic, in contrast to the nonmonotonic and symmetric structure observed before.
For the physically accessible state $\rho_W^{AB_{\text{out}}}$, 
$F_{d}(\rho_W^{AB_{\text{out}}})$ decreases monotonically with increasing $\delta$ at fixed $p$, indicating that the Hawking radiation degrades the efficiency of imaginarity distillation. 
In contrast, for the physically inaccessible state $\rho_W^{AB_{\text{in}}}$, 
$F_{d}(\rho_W^{AB_{\text{in}}})$ increases monotonically with $\delta$, 
indicating a corresponding enhancement of distillation capability.
The maximal value of $F_{d}(\rho_W^{AB_{\text{in}}})$ is achieved at $p=1$ in the high-temperature limit $\delta \to \pi/4^{-}$, 
where $F_{d} \to \frac{1}{2}(1+1/\sqrt{2}) \approx 0.8536$. 
For the special case $p=0$, the fidelity remains $F_{d}=1/2$ independently of $\delta$, 
while at $\delta=0$, one has $F_{d}(\rho_W^{AB_{\text{in}}})=1/2$ for all $p$.

\section{Conclusion}
\label{sec:Conclusion}
In this work, we investigated quantum imaginarity in Schwarzschild spacetime through two operational protocols, corresponding to NAQI and assisted imaginarity distillation.
The key quantities characterizing these tasks, the NAQI gap and the assisted fidelity of imaginarity, describe how imaginarity is affected by Hawking radiation in physically accessible and inaccessible regions.

In the NAQI scenario, we characterized the behavior of the NAQI gap using both the $l_1$ norm and the relative entropy measures of imaginarity.
Our results demonstrate that Hawking radiation induces a clear distinction between physically accessible and inaccessible modes.
In the physically accessible region, the NAQI gap decreases monotonically with increasing Hawking temperature.
In particular, the gap can approach zero and become negative, indicating that the positivity condition required for NAQI is satisfied only within a restricted parameter regime.
In this regime, when the NAQI condition is met, the corresponding states exhibit steerability.
In contrast, the physically inaccessible region exhibits an opposite thermal dependence, where the NAQI gap remains negative throughout, precluding the activation of NAQI in this sector.
Moreover, the behavior of the NAQI gap depends sensitively on the structure of the initial mixed state and its interplay with Hawking-induced effects.

In parallel, we studied assisted imaginarity distillation under the same Hawking-induced dynamics. 
The assisted fidelity exhibits a similar temperature dependence between the two regions, decreasing in the physically accessible region while increasing in the physically inaccessible region. 
This leads to a reduction of distillation capability in the accessible sector and an enhancement in the inaccessible sector.

Overall, our results demonstrate that Hawking radiation induces a temperature-dependent modification of quantum imaginarity in Schwarzschild spacetime.
Although arising from distinct operational tasks, the NAQI gap and assisted fidelity exhibit similar trends under temperature variation.
This shows that Hawking radiation affects both the emergence of NAQI and the efficiency of imaginarity distillation.
Relativistic effects therefore change how quantum imaginarity behaves as a resource. 
These results may motivate further studies of quantum resource theories in relativistic quantum information settings.

\section{Acknowledgement}
This work was supported by the Startup Funding of Guangdong Polytechnic Normal University (Grant No. 2021SDKYA178), Key Laboratory of Computational Science and the Application of Hainan Province (Grant No. JSKX202503), the Research Fund of Jianghan University (Grant No. 2023JCYJ08), in part by the National Natural Science Foundation of China (NSFC) (Grant No. 12301582), GDSTA( SKXRC2025442), the
China Scholarship Council (CSC), and the Natural Science Foundation
of Hainan Province (Grant No. 125RC744).

\bibliography{Bib}
\newpage
\appendix

\section{Projective Measurements and Mutually Unbiased Bases}
For a two-qubit  state $\rho^{AB}$, Alice performs three binary projective measurements $\mathcal{P}=\{P_x\}_{x=1}^3$, where each setting $x$ is associated with projectors $\{P_{o|x}\}_{o=0,1}$ defined as
\begin{align}
P_{o|x} = |\psi_{o|x}\rangle\langle \psi_{o|x}|,
\end{align}
with
\begin{align*}
|\psi_{0|x}\rangle &= \cos \frac{\theta_x}{2} |0\rangle + e^{\mathrm{i}\phi_x} \sin \frac{\theta_x}{2} |1\rangle, \\
|\psi_{1|x}\rangle &= \sin \frac{\theta_x}{2} |0\rangle - e^{\mathrm{i}\phi_x} \cos \frac{\theta_x}{2} |1\rangle,
\end{align*} 
where $\theta_{x}\in[0,\pi]$ and $\phi_{x}\in[0,2\pi)$.
Conditioned on outcome $o$ of setting $x$, Bob obtains the (unnormalized) conditional states
$\rho^B_{o|x}=\operatorname{Tr}_A\left[(P_{{o|x}}\otimes \1)\rho^{AB}\right]$ with probability $p(o|x)=\operatorname{Tr}\left[(P_{o|x}\otimes \1)\rho^{AB}\right]$, forming the ensemble
$\mathcal{E}_x=\{p(o|x), \rho^B_{o|x}\}$. 
We then evaluate the imaginarity of Bob’s states with respect to a parametrized family of mutually unbiased bases  $\mathcal{B}=\{\mathcal{B}_x\}_{x=1}^3$, defined as 
\begin{align}
\{|\mathcal{B}_1^{\pm} \rangle\} &= \{ \cos \frac{\theta_4}{2} |0\rangle + e^{\mathrm{i}\phi_4} \sin \frac{\theta_4}{2} |1\rangle,\notag\\
&\quad\quad\sin \frac{\theta_4}{2} |0\rangle - e^{\mathrm{i}\phi_4} \cos \frac{\theta_4}{2} |1\rangle \},\\
\{ |\mathcal{B}_2^{\pm} \rangle\} &= \{ \frac{|\mathcal{B}_1^{+}\rangle \pm |\mathcal{B}_1^{-}\rangle}{\sqrt{2}} \},\\
\{ |\mathcal{B}_3^{\pm} \rangle\} &= \{ \frac{|\mathcal{B}_1^{+}\rangle \pm \mathrm{i} |\mathcal{B}_1^{-}\rangle}{\sqrt{2}} \}
\end{align}
where $\theta_4 \in [0, \pi]$ and $\phi_4 \in [0, 2\pi)$.
In the qubit case, any two complete sets of three mutually unbiased bases are related by a unitary transformation, so this parametrization is without loss of generality.
The measurement strategy $\mathcal{P}$ together with the basis set $\mathcal{B}$ is parametrized by eight real parameters $\{\theta_x, \phi_x\}_{x=1}^3$ and $\{\theta_4, \phi_4\}$.
We adopt this parametrization as a numerical ansatz for exploring projective measurements and mutually unbiased bases in Eq.~(\ref{optim}).

\section{Quantum State Evolution in Schwarzschild Spacetime}
Any two-qubit state $\rho^{AB}$ admits the Bloch representation
\begin{align}
\rho=\frac{1}{4} ( \1 \otimes \1 + \vec{a} \cdot \vec{\sigma} \otimes \1 + \1 \otimes \vec{b} \cdot \vec{\sigma} + \sum_{i,j=1}^{3} E_{ij} \, \sigma_i \otimes \sigma_j ),
\end{align}
where $\1$ denotes the identity matrix, $\{\sigma_i\}_{i=1}^3$ are Pauli matrices, and $\vec{\sigma} = (\sigma_1, \sigma_2, \sigma_3)$ is the Pauli vector.
The local Bloch vectors $\vec{a} = (a_1, a_2, a_3)$ and $\vec{b} = (b_1, b_2, b_3)$ are defined as  $a_i = \operatorname{Tr}[\rho \, (\sigma_i \otimes \1)]$ and $b_j = \operatorname{Tr}[\rho \, (\1 \otimes \sigma_j)]$, respectively.
The correlation matrix elements are given by $E_{ij}=\operatorname{Tr}[\rho (\sigma_i \otimes \sigma_j) ]$.
With this framework, the Bell-diagonal mixed state in Eq.~(\ref{eg1}) and Werner state in Eq.~(\ref{eg2}) $\rho^{AB}_W$ can be expressed as
\begin{align}
\rho^{AB}=&\frac{1}{4}
\big(\1 \otimes \1+  \sigma_1\otimes \sigma_1 + (1-2p)\sigma_2 \otimes \sigma_2\notag\\
&+ (2p-1)\sigma_3 \otimes \sigma_3\big), 
\end{align}
and
\begin{align}
 \rho^{AB}_W=&\frac{1}{4}\big(\1 \otimes \1+p   \sigma_1\otimes \sigma_1 -p \sigma_2 \otimes \sigma_2\notag\\
&+ p\sigma_3 \otimes \sigma_3\big).  
\end{align}

In the presence of Hawking radiation, the field modes across the event horizon undergo a decomposition into exterior and interior sectors.
Following  Ref.~\cite{discord_xiong2026}, tracing out one sector induces an effective Bloch representation of the Pauli operators, which we use directly in the subsequent calculations.
After tracing out the interior modes, the Pauli operators associated with the exterior sector are given by
\begin{align}
(\1)_{\text{out}}&=\1-\frac{1}{e^{\frac{\omega}{T}}+1}\sigma_3, \quad (\sigma_1)_{\text{out}}=\frac{1}{\sqrt{e^{-\frac{\omega}{T}}+1}}\sigma_1\notag\\(\sigma_2)_{\text{out}}&=\frac{1}{\sqrt{e^{-\frac{\omega}{T}}+1}}\sigma_2,\quad (\sigma_3)_{\text{out}}=\frac{1}{e^{-\frac{\omega}{T}}+1}\sigma_3,
\end{align}
Similarly, tracing out the exterior modes leads to the corresponding effective representation in the interior sector,
\begin{align}
(\1)_{\text{in}}&=\1+\frac{1}{e^{-\frac{\omega}{T}}+1}\sigma_3, \quad (\sigma_1)_{\text{in}}=\frac{1}{\sqrt{e^{\frac{\omega}{T}}+1}}\sigma_1\notag\\(\sigma_2)_{\text{in}}&=-\frac{1}{\sqrt{e^{\frac{\omega}{T}}+1}}\sigma_2,\quad (\sigma_3)_{\text{in}}=-\frac{1}{e^{\frac{\omega}{T}}+1}\sigma_3.
\end{align}
Consequently, for a bipartite state $\rho^{AB}$ with Alice in the asymptotically flat region and Bob near the event horizon, the reduced states $\rho^{AB_{\text{out}}}$ and $\rho^{AB_{\text{in}}}$ can be obtained by expressing the state in terms of the above operator representation.
The same procedure applies to the Werner state $\rho_W^{AB}$, yielding $\rho_W^{AB_{\text{out}}}$ and $\rho_W^{AB_{\text{in}}}$.

\end{document}